\documentclass[runningheads]{llncs}
\usepackage[T1]{fontenc}
\usepackage{graphicx,verbatim}
\usepackage{amsmath,amssymb}
\usepackage{booktabs}
\usepackage{multirow}
\usepackage{xcolor}
\usepackage{enumitem}

\begin{document}





\title{Clinical-Injection Transformer with Domain-Adapted MAE for\\ Lupus Nephritis Prognosis Prediction}

\author{Yuewen Huang\inst{1} \and
Zhitao Ye\inst{2} \and
Guangnan Feng\inst{1} \and
Fudan Zheng\inst{1}\thanks{Corresponding authors.} \and
Xia Gao\inst{2}\textsuperscript{*} \and
Yutong Lu\inst{1}\textsuperscript{*}}

\authorrunning{Y. Huang et al.}

\institute{Sun Yat-sen University \and
Department of Nephrology, Guangzhou Women and Children's Medical Center, Guangzhou Medical University, Guangzhou, China \\
\email{huangyw66@mail2.sysu.edu.cn, yezht@foxmail.com, fenggn7@mail.sysu.edu.cn, zhengfd5@mail.sysu.edu.cn, gaoxiagz@vip.163.com, luyutong@mail.sysu.edu.cn}}

\maketitle

\begin{abstract}
Lupus nephritis (LN) is a severe complication of systemic lupus erythematosus that affects pediatric patients with significantly greater severity and worse renal outcomes compared to adults. Despite the urgent clinical need, predicting pediatric LN prognosis remains unexplored in computational pathology. Furthermore, the only existing histopathology-based approach for LN relies on multiple costly staining protocols and fails to integrate complementary clinical data. To address these gaps, we propose the first multimodal computational pathology framework for three-class treatment response prediction (complete remission, partial response, and no response) in pediatric LN, utilizing only routine PAS-stained biopsies and structured clinical data. Our framework introduces two key methodological innovations. First, a Clinical-Injection Transformer (CIT) embeds clinical features as condition tokens into patch-level self-attention, facilitating implicit and bidirectional cross-modal interactions within a unified attention space. Second, we design a decoupled representation-knowledge adaptation strategy using a domain-adapted Masked Autoencoder (MAE). This strategy explicitly separates self-supervised morphological feature learning from pathological knowledge extraction. Additionally, we introduce a multi-granularity morphological type injection mechanism to bridge distilled classification knowledge with downstream prognostic predictions at both the instance and patient levels. Evaluated on a cohort of 71 pediatric LN patients with KDIGO-standardized labels, our method achieves a three-class accuracy of 90.1\% and an AUC of 89.4\%, demonstrating its potential as a highly accurate and cost-effective prognostic tool.

\end{abstract}

\section{Introduction}

Lupus nephritis (LN) affects 50--82\% of pediatric systemic lupus erythematosus (SLE) patients~\cite{ref_sinha_bagga}---substantially more than adults (20--40\%)~\cite{ref_ln_epidemiology}---with greater disease severity and worse long-term renal outcomes~\cite{ref_chan_cln}. 
Yet childhood-onset SLE itself is exceptionally rare, with an incidence of only 0.3--0.9 per 100,000 children-years---approximately one-sixth of the adult rate~\cite{ref_kamphuis,ref_tian_sle}. When further restricted to biopsy-confirmed cases with longitudinal follow-up and digitized pathology, even the largest multicenter pediatric LN cohorts comprise only about 300 patients~\cite{ref_chan_cln}. 
Besides, its treatment response varies significantly---complete remission (CR), partial response (PR), or no response (NR)---and early identification is critical. 

With the growing adoption of deep learning in medical image analysis, existing approaches for LN prognosis fall into two disjoint tracks: machine learning models based on clinical biomarkers have shown promise for treatment outcome and relapse prediction~\cite{ref_ln_flare_dl}, but discard the rich morphological information in renal biopsies; on the other hand, the only histopathology-based approach~\cite{ref_cheng_ki2024} requires four costly staining protocols without clinical data integration. No method combines histopathology with clinical data for LN prognosis, and pediatric LN~\cite{ref_sinha_bagga} lacks any biopsy-image-based prediction approach.

Meanwhile, multimodal fusion of histopathology and clinical or genomic data has emerged as a key direction. 
However, these methods such as MCAT~\cite{ref_mcat}, SurvPath~\cite{ref_survpath}, CMTA~\cite{ref_cmta}, and HEALNet~\cite{ref_healnet}, typically require complex dual-stream architectures designed for large-scale datasets, making them prone to overfitting on small cohorts typical of rare pediatric diseases. This motivates our Clinical-Injection Transformer, which injects clinical features as condition tokens into a unified self-attention space for parameter-efficient cross-modal interaction.

In addition, self-supervised pretraining, such as Masked Autoencoder (MAE)-based approaches~\cite{ref_mae} have shown effectiveness for medical image classification~\cite{ref_selfmedmae} and pathology representation learning~\cite{ref_pama}. These motivate us to explore MAE-based self-supervised pretraining on glomerulus patches to learn domain-specific representations.

In this work, we propose a multimodal framework for pediatric LN three-class prognosis prediction. Our contributions are:
\begin{enumerate}[leftmargin=*,itemsep=1pt,topsep=2pt]
\item The \textbf{first multimodal computational pathology framework} for LN treatment response prediction from routine single-stain histopathology with clinical data, under KDIGO-standardized three-class criteria (CR/PR/NR).
\item \textbf{Clinical-Injection Transformer (CIT)}, which injects clinical features as condition tokens into patch-level self-attention, enabling bidirectional cross-modal interaction---simpler and more parameter-efficient than cross-attention~\cite{ref_mcat} or separate fusion modules~\cite{ref_survpath}.
\item \textbf{Decoupled representation-knowledge adaptation} that separates self-supervised feature learning from morphological classification, preserving prognostically relevant features (+7.1\% over DINOv2~\cite{ref_dinov2}).
\item \textbf{Multi-granularity morphological type injection} bridging distilled classification knowledge with prognosis at patch and patient levels (+2.3\% Acc, +5.3\% M-F1).
\end{enumerate}

\section{Method}

\subsection{Framework Overview}

\begin{figure}[t]
\centering
\includegraphics[width=1\textwidth]{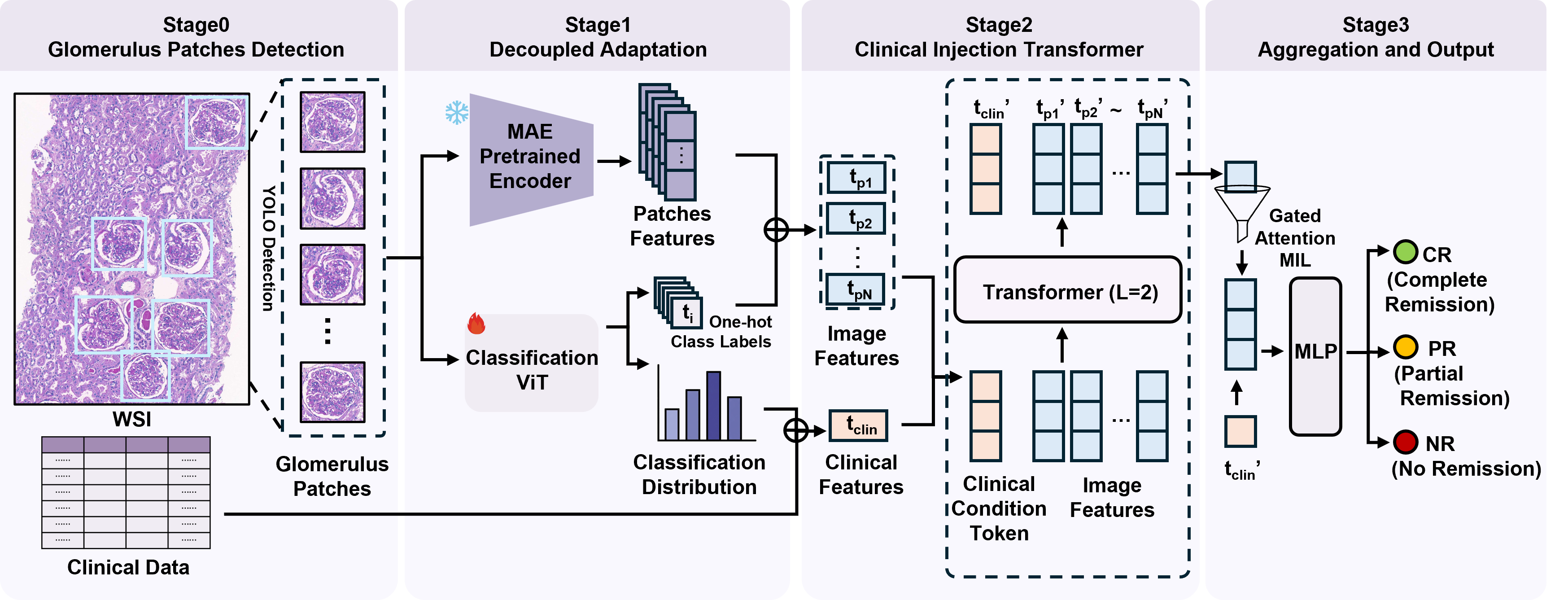}
\caption{Overview of the proposed framework}
\label{fig:pipeline}
\end{figure}

Our framework (Fig.~\ref{fig:pipeline}) consists of four stages: (0) automated glomerulus detection from PAS-stained WSIs, producing $N$ cropped glomerulus patches per patient; (1) decoupled representation-knowledge adaptation---a ViT-B/16 encoder is pretrained via MAE~\cite{ref_mae} on 4,826 glomerulus patches, then serves two paths: the \emph{frozen} pretrained encoder extracts patch representations ${\mathbf{x}i \in \mathbb{R}^{768}}{i=1}^{N}$ (representation path), while a \emph{finetuned} copy with a classification head produces discrete morphological type labels ${t_i}_{i=1}^{N}$ (knowledge path), which are combined with clinical features $\mathbf{c} \in \mathbb{R}^{d_c}$; (2) Clinical-Injection Transformer fuses patch features (with multi-granularity type injection) and clinical condition token via unified self-attention for bidirectional cross-modal interaction; (3) gated attention-based multiple instance learning (MIL) aggregation pools enriched patch tokens into a patient-level representation, which is concatenated with the enriched clinical token and classified into CR/PR/NR via a classification head.

\subsection{Decoupled Representation-Knowledge Adaptation}

Finetuning for morphological classification narrows the representation space toward category-level discrimination, at the cost of subtle morphological cues---such as texture variations and structural irregularities---that are not rewarded by the classification objective yet may prove essential for predicting treatment response. We address this by \emph{decoupling} domain adaptation into two complementary paths:

\textbf{Representation path: self-supervised pretraining.} A ViT-B/16 encoder is pretrained using masked autoencoding~\cite{ref_mae} on approximately 5,000 glomerulus patches, including publicly available annotated glomerular data~\cite{ref_holohisto} with an in-house collection from 137 patients at a collaborating center. This expanded pretraining corpus ensures exposure to diverse glomerular morphologies across institutions and staining variations. Following standard MAE, 75\% of patches are masked and the encoder learns to reconstruct the masked regions, capturing broad tissue textures and structural patterns. The pretrained encoder is \emph{frozen} for downstream feature extraction, preserving its rich, task-agnostic representations.

\textbf{Knowledge path: supervised morphological classification.} A copy of the pretrained encoder is finetuned for 5-class glomerular morphological classification (achieving 92\% accuracy): mesangial proliferative, normal, endocapillary proliferative, crescentic, and sclerotic.
Rather than using the finetuned encoder for feature extraction, we \emph{distill} its acquired pathological knowledge into discrete type labels, which are subsequently injected into the downstream model via multi-granularity type injection (Sec.~2.4).

Our ablation (Sec.~3.2) confirms this effect: frozen self-supervised features outperform finetuned features by +6.0\% Acc, indicating that classification-oriented adaptation indeed discards prognostically relevant information. By separating representation learning from knowledge extraction, we preserve the representational richness of self-supervised features while still leveraging pathological semantics through structured knowledge injection.

\subsection{Clinical-Injection Transformer (CIT)}

The core of our fusion module integrates clinical features with patch-level image features through a unified self-attention mechanism.

\textbf{Clinical condition token.} Given clinical features $\mathbf{c} \in \mathbb{R}^{d_c}$, we project them into the patch feature space:
\begin{equation}
\mathbf{z}_{\text{cond}} = \text{LayerNorm}(\text{ReLU}(\mathbf{W}_c \mathbf{c} + \mathbf{b}_c)) \in \mathbb{R}^{d_h}
\end{equation}
where $d_h{=}256$ is the hidden dimension. This condition token represents the patient's clinical context.

\textbf{Unified self-attention.} Patch features $\{\mathbf{x}_i\}_{i=1}^N$ are projected to $\{\mathbf{z}_i \in \mathbb{R}^{d_h}\}$ and concatenated with the condition token:
\begin{equation}
\mathbf{Z}^{(0)} = [\mathbf{z}_{\text{cond}}; \mathbf{z}_1; \mathbf{z}_2; \ldots; \mathbf{z}_N] \in \mathbb{R}^{(N+1) \times d_h}
\end{equation}
This sequence is processed by $L{=}2$ Transformer encoder layers (4 attention heads per layer):
\begin{equation}
\mathbf{Z}^{(l)} = \text{TransformerLayer}(\mathbf{Z}^{(l-1)}), \quad l = 1, \ldots, L
\end{equation}
By placing clinical and image tokens in the \emph{same} self-attention sequence, each clinical feature naturally attends to all patch features and vice versa, enabling \textbf{implicit bidirectional interaction} without explicit cross-attention modules. The entire CIT module contains approximately 0.56M parameters, substantially fewer than dual-stream cross-attention architectures, mitigating overfitting on small datasets.

\textbf{Attention-based MIL aggregation.} After the Transformer, patch representations are aggregated into a patient-level image representation via gated attention-based MIL pooling~\cite{ref_abmil}:
\begin{equation}
a_i = \frac{\exp(\mathbf{w}^\top (\tanh(\mathbf{V}\mathbf{z}_i^{(L)}) \odot \sigma(\mathbf{U}\mathbf{z}_i^{(L)})))}{\sum_j \exp(\mathbf{w}^\top (\tanh(\mathbf{V}\mathbf{z}_j^{(L)}) \odot \sigma(\mathbf{U}\mathbf{z}_j^{(L)})))}, \quad \mathbf{h}_{\text{img}} = \sum_{i=1}^{N} a_i \, \mathbf{z}_i^{(L)}
\end{equation}
The final patient representation $\mathbf{h} = [\mathbf{h}_{\text{img}}; \mathbf{z}_{\text{cond}}^{(L)}]$ concatenates the aggregated image representation with the enriched clinical token, followed by a classification head.

\subsection{Multi-Granularity Morphological Type Injection}

We bridge the knowledge path and downstream prognosis through multi-granularity feature injection, providing the model with explicit morphological semantics at complementary hierarchical levels:

\textbf{Patch-level injection.} Each patch's morphological type label (from the knowledge path) is encoded as a one-hot vector $\mathbf{t}_i \in \{0,1\}^K$ ($K$=5 classes) and concatenated with the image feature: $\tilde{\mathbf{x}}_i = [\mathbf{x}_i; \mathbf{t}_i] \in \mathbb{R}^{d+K}$. This informs the Transformer of each patch's morphological identity.

\textbf{Patient-level injection.} The type distribution $\mathbf{d} = \frac{1}{N}\sum_i \mathbf{t}_i \in [0,1]^K$ is concatenated with clinical features: $\tilde{\mathbf{c}} = [\mathbf{c}; \mathbf{d}] \in \mathbb{R}^{d_c+K}$, encoding overall lesion composition (e.g., proportion of sclerotic or crescentic glomeruli).

Unlike multi-granularity approaches operating within a single modality~\cite{ref_crossscale_mil}, our injection spans \emph{across} modalities, transferring distilled classification knowledge to prognosis at corresponding hierarchical levels. Patch-level morphological identity constrains instance-level attention, while patient-level composition informs global clinical interaction; these two granularities are designed to act jointly and provide complementary signals. To regularize training under severe class imbalance, we apply manifold Mixup~\cite{ref_manifold_mixup} at the patient-level representation space, where interpolation in the learned embedding captures semantically meaningful morphological summaries.

\section{Experiments}

\subsection{Dataset and Setup}


We collected PAS-stained renal biopsy WSIs and clinical records from 180 pediatric LN patients at a tertiary children's hospital. A YOLO-based detection model~\cite{ref_holohisto} extracts glomeruli (99\% sensitivity). Clinical features include demographics, laboratory values, and ISN/RPS classification (25 dimensions at baseline; 59 with 3-month delta features), imputed via MICE~\cite{ref_mice}. After inclusion criteria, 71 patients (CR=49, PR=10, NR=12) with 2,925 patches (448$\times$448, mean 31.5/patient) formed the cohort. The MAE pretraining corpus includes patches from the study cohort. 

KDIGO labels~\cite{ref_kdigo}: CR requires proteinuria normalization and stable creatinine; PR requires $\geq$50\% proteinuria reduction; NR otherwise. We evaluate baseline-only (0m) and baseline+3-month (0m+3m) configurations. All experiments use 5-fold cross-validation with 3 seeds, AdamW (lr=1e-3, wd=5e-4), cosine annealing, early stopping (patience 50), weighted cross-entropy, manifold Mixup~\cite{ref_manifold_mixup} ($\alpha$=0.4), and label smoothing ($\epsilon$=0.05).

\subsection{Main Results}

Table~\ref{tab:main} presents three-class prognosis prediction results. We evaluate the progression from single-modality prediction to multimodal fusion with increasing clinical context. ABMIL~\cite{ref_abmil}, TransMIL~\cite{ref_transmil}, and CLAM~\cite{ref_clam} serve as image-only MIL baselines, Clinical MLP provides the single-modality clinical baseline, and Late Fusion concatenates image and clinical features without cross-modal interaction. Our CIT is evaluated with two clinical input configurations: baseline-only (0m, 25 dimensions) and baseline plus 3-month follow-up (0m+3m, 59 dimensions including treatment response features).

\begin{figure}[t]
\centering
\includegraphics[width=0.75\textwidth]{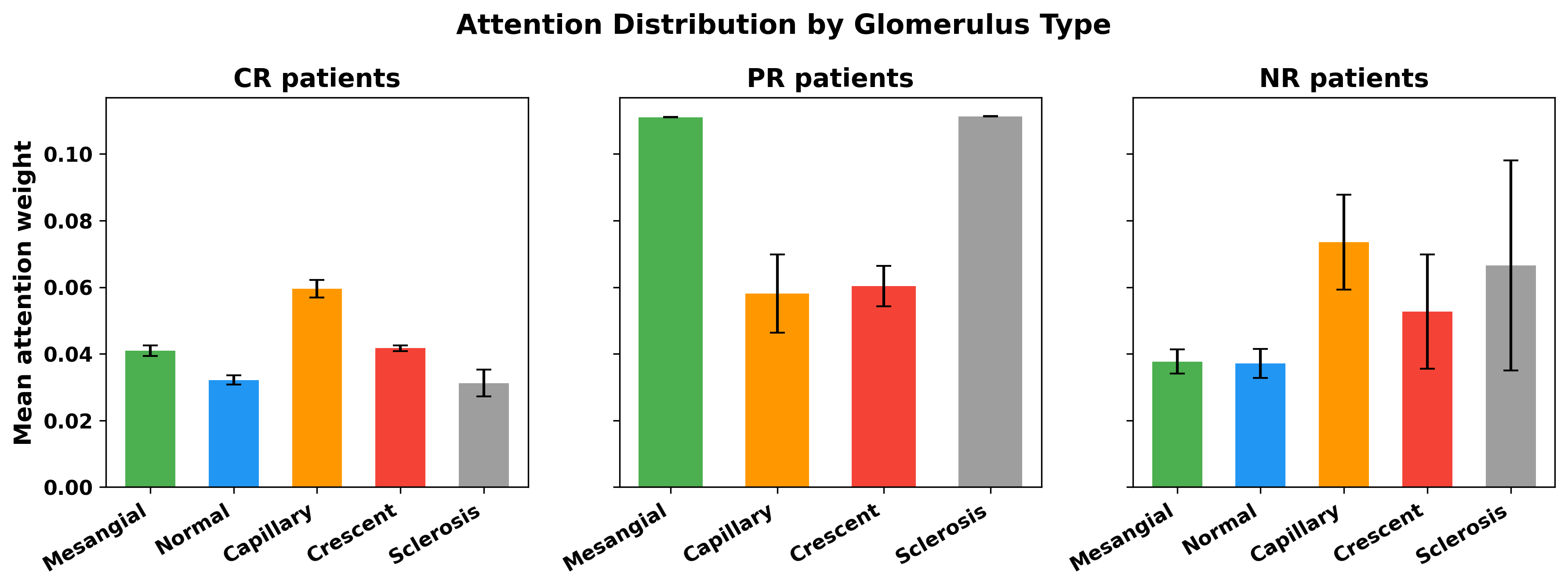}
\caption{Mean MIL attention weights by glomerulus morphological type for each outcome class (3-seed aggregate, 0m+3m$\rightarrow$12m). PR and NR patients show elevated attention to pathological types (mesangial proliferative, sclerotic), while CR patients exhibit more uniform attention distribution.}
\label{fig:analysis}
\end{figure}

Key observations: (1) \textbf{Single modalities are insufficient}: all image-only MIL baselines (ABMIL 68.6\%, TransMIL 67.6\%, CLAM 65.8\% Acc) plateau with low M-F1 ($\leq$51.3\%), indicating poor minority class discrimination regardless of attention architecture. 
Clinical MLP (80.7\% Acc) performs substantially better with temporal features but remains insufficient for reliable three-class discrimination.
(2) \textbf{Cross-modal fusion provides substantial improvement}: Late Fusion combining modalities through simple concatenation reaches 83.1\% Acc, and CIT's conditional attention mechanism further provides more effective cross-modal interaction, reaching 90.1\% Acc (+7.0\% over Late Fusion). (3) \textbf{Temporal follow-up enhances prediction}: incorporating 3-month features improves CIT from 86.4\% to 90.1\% Acc, with notable gains in performance in minority class (M-F1: 77.0\%$\rightarrow$83.9\%) and overall discrimination (AUC: 88.0\%$\rightarrow$89.4\%). 
The 3-month timepoint, a clinically established assessment point per KDIGO guidelines~\cite{ref_kdigo}, captures early treatment response dynamics while the 12-month outcome remains uncertain. By integrating baseline and 3-month features, our model achieves prognostic predictions 6 months prior to the final outcome, providing a clinically actionable window for timely therapeutic adjustments. This extended prediction horizon underscores the effectiveness of our approach in leveraging early response signals to forecast long-term trajectories. (4) \textbf{Attention maps reveal clinically meaningful patterns}: The MIL attention distribution (Fig.~\ref{fig:analysis}) reveals that PR and NR patients show elevated attention to mesangial proliferative and sclerotic glomeruli---pathological types associated with disease activity and chronicity---while CR patients exhibit more uniform attention across types, reflecting their milder histological profiles. This suggests the model learns to focus on morphologically abnormal glomeruli for adverse outcome prediction. All improvements of CIT over baselines are statistically significant (Wilcoxon signed-rank test on 15 paired fold-level observations).

\begin{table}[t]
\caption{Three-class prognosis prediction (CR/PR/NR) on $n$=71 patients. Results are 5-fold cross-validation mean$\pm$std. Best in \textbf{bold}.}\label{tab:main}
\centering
\fontsize{8}{9.5}\selectfont
\begin{tabular}{@{}lcccc@{}}
\toprule
Method & Acc(\%) & W-F1(\%) & M-F1(\%) & AUC(\%) \\
\midrule
ABMIL~\cite{ref_abmil} (image-only) & 68.6$\pm$1.8 & 62.1$\pm$1.0 & 38.9$\pm$2.0 & 45.2$\pm$3.3 \\
TransMIL~\cite{ref_transmil} (image-only) & 67.6$\pm$2.1 & 63.5$\pm$2.1 & 43.5$\pm$2.1 & 60.6$\pm$0.6 \\
CLAM~\cite{ref_clam} (image-only) & 65.8$\pm$0.8 & 65.4$\pm$1.4 & 51.3$\pm$2.5 & 61.0$\pm$1.5 \\
Clinical MLP (0m+3m) & 80.7$\pm$4.3 & 80.4$\pm$3.3 & 70.2$\pm$4.9 & 81.9$\pm$3.2 \\
Late Fusion (0m+3m) & 83.1$\pm$2.3 & 82.5$\pm$2.7 & 74.6$\pm$3.4 & 85.4$\pm$2.1 \\
CIT (Ours, 0m) & 86.4$\pm$2.6 & 85.0$\pm$2.6 & 77.0$\pm$4.3 & 88.0$\pm$1.9 \\
CIT (Ours, 0m+3m) & \textbf{90.1$\pm$1.1} & \textbf{89.2$\pm$1.0} & \textbf{83.9$\pm$1.3} & \textbf{89.4$\pm$2.9} \\
\bottomrule
\end{tabular}
\end{table}

\subsection{Ablation Studies}

We conduct systematic ablation experiments to validate each component's contribution. All ablations are performed on the main 0m+3m$\rightarrow$12m setting ($n$=71) for consistency.

\textbf{Fusion architecture} (Table~\ref{tab:ablation}a). We compare CIT against cross-attention fusion~\cite{ref_mcat} and late fusion.
 CIT outperforms cross-attention by +7.0\% accuracy (90.1\% vs. 83.1\%) and late fusion by +7.0\% accuracy (90.1\% vs. 83.1\%), demonstrating that condition-token injection is more effective than separate cross-attention modules or independent-branch fusion on small datasets.

\begin{table}[t]
\caption{Ablation studies. (a) Fusion architecture, (b) feature extractor, and (c) morphological type injection, all on 0m+3m$\rightarrow$12m ($n$=71). 5-fold cross-validation, mean$\pm$std over 3 seeds.}\label{tab:ablation}
\centering
\fontsize{8}{9.5}\selectfont
\begin{tabular}{@{}lcccc@{}}
\toprule
\multicolumn{5}{l}{\textbf{(a) Fusion Architecture} (0m+3m$\rightarrow$12m, $n$=71)} \\
\midrule
Method & Acc(\%) & W-F1(\%) & M-F1(\%) & AUC(\%) \\
\midrule
Cross-Attention~\cite{ref_mcat} & 83.1$\pm$3.9 & 83.6$\pm$3.7 & 76.7$\pm$5.8 & 87.1$\pm$5.5 \\
Late Fusion & 83.1$\pm$2.3 & 82.5$\pm$2.7 & 74.6$\pm$3.4 & 85.4$\pm$2.1 \\
CIT (Ours) & \textbf{90.1$\pm$1.1} & \textbf{89.2$\pm$1.0} & \textbf{83.9$\pm$1.3} & \textbf{89.4$\pm$2.9} \\
\midrule
\multicolumn{5}{l}{\textbf{(b) Feature Extractor} (0m+3m$\rightarrow$12m, $n$=71)} \\
\midrule
Feature Source & Acc(\%) & W-F1(\%) & M-F1(\%) & AUC(\%) \\
\midrule
DINOv2 ViT-B/14~\cite{ref_dinov2} & 83.0$\pm$4.0 & 82.3$\pm$4.1 & 75.0$\pm$5.3 & 83.1$\pm$5.3 \\
ResNet50 (ImageNet) & 82.0$\pm$1.1 & 81.5$\pm$1.2 & 72.7$\pm$1.6 & 81.4$\pm$2.6 \\
MAE ViT-B/16 (finetuned) & 84.1$\pm$2.4 & 84.1$\pm$2.2 & 76.9$\pm$3.7 & 83.0$\pm$3.0 \\
MAE ViT-B/16 (pretrained, Ours) & \textbf{90.1$\pm$1.1} & \textbf{89.2$\pm$1.0} & \textbf{83.9$\pm$1.3} & \textbf{89.4$\pm$2.9} \\
\midrule
\multicolumn{5}{l}{\textbf{(c) Morphological Type Injection} (0m+3m$\rightarrow$12m, $n$=71)} \\
\midrule
Configuration & Acc(\%) & W-F1(\%) & M-F1(\%) & AUC(\%) \\
\midrule
No type features & 87.8$\pm$2.4 & 86.5$\pm$2.6 & 78.4$\pm$3.2 & 85.4$\pm$5.2 \\
Multi-granularity injection (Ours) & \textbf{90.1$\pm$1.1} & \textbf{89.2$\pm$1.0} & \textbf{83.9$\pm$1.3} & \textbf{89.4$\pm$2.9} \\
\bottomrule
\end{tabular}
\end{table}

\textbf{Feature extractor} (Table~\ref{tab:ablation}b). Domain-adapted MAE pretrained features outperform DINOv2 by +7.1\% Acc and ResNet50 by +8.1\%. MAE finetuned features (84.1\%) show only marginal gains over these baselines, confirming that finetuning narrows representations while frozen self-supervised features preserve richer morphological diversity.

\textbf{Morphological type injection} (Table~\ref{tab:ablation}c). In this ablation, we evaluate the contribution of our proposed multi-granularity injection of glomerular morphological type labels (Sec.~2.4), where type information is simultaneously provided at both the patch and patient levels. This design leverages the complementary nature of morphological cues across scales: patch-level features constrain local instance attention, while patient-level composition informs global clinical context. The multi-granularity injection achieves the best performance across all four metrics, with 90.1\% accuracy (+2.3\% over the no-type baseline of 87.8\%), 83.9\% macro F1 (+5.5\%), and 89.4\% AUC (+4.0\%). 


\section{Conclusion}

We presented a multimodal computational pathology framework for pediatric lupus nephritis three-class prognosis prediction, integrating automated glomerulus detection, domain-adapted MAE feature learning, and a Clinical-Injection Transformer for multimodal fusion. 
Our CIT enables efficient bidirectional cross-modal interaction through condition token injection, while decoupled representation-knowledge adaptation preserves prognostically relevant features and leverages morphological knowledge through multi-granularity type injection. On 71 patients, our method achieves 90.1\% accuracy and 89.4\% AUC using KDIGO-standardized labels, demonstrating AI-assisted prognosis potential in rare pediatric kidney diseases.

Limitations include the single-center design and modest cohort size.
However, this scale is consistent with the state of the field—even the largest multicenter studies of biopsy-confirmed childhood lupus nephritis with long-term follow-up report approximately 300 patients~\cite{ref_chan_cln}. This reflects both the low incidence of pediatric SLE (0.3–0.9 per 100,000 children-years) and the compounded challenge of acquiring invasive renal biopsies, digitizing histopathology, and maintaining longitudinal follow-up. To address these constraints, we are pursuing multi-center validation and integration of longitudinal imaging data in ongoing work.

\end{document}